\documentclass{emulateapj}

\newcommand{\mbfB}{\mathbf{B}}
\newcommand{\mbfu}{\mathbf{u}}

\newcommand{\f}   {\frac}

\newcommand{\lomega}{\lambda_\Omega}

\begin{document}

\title{Magnetized Non-linear Thin Shell Instability: Numerical Studies in 2D}

\author{Fabian Heitsch\altaffilmark{1}}
\author{Adrianne D. Slyz\altaffilmark{2,3}}
\author{Julien E.G. Devriendt\altaffilmark{2}}
\author{Lee W. Hartmann\altaffilmark{1}}
\author{Andreas Burkert\altaffilmark{4}}
\altaffiltext{1}{Dept. of Astronomy, University of Michigan, 500 Church St., 
                 Ann Arbor, MI 48109-1042, U.S.A}
\altaffiltext{2}{Universit\'e Claude Bernard Lyon 1, 
                CRAL, Observatoire de Lyon, 9 Avenue Charles Andr\'{e},
                 69561 St-Genis Laval Cedex, France; CNRS, UMR 5574; ENS Lyon}
\altaffiltext{3}{Oxford University, Astrophysics, Denys Wilkinson Building, Keble Road,
                 Oxford, OX1 3RH, United Kingdom}
\altaffiltext{4}{University Observatory Munich, Scheinerstr. 1, 81679 Munich, Germany}
\lefthead{Heitsch et al.}
\righthead{Magnetized NTSI in 2D}

\begin{abstract}
We revisit the analysis of the Non-linear Thin Shell Instability (NTSI)
numerically, including magnetic fields. The magnetic tension force is expected to 
work against the main driver of the NTSI -- namely transverse momentum transport. 
However, depending on the field strength and orientation,
the instability may grow. For fields aligned with the inflow, we find that the NTSI is suppressed
only when the Alfv\'en speed surpasses the (supersonic) velocities
generated along the collision interface. Even for fields perpendicular 
to the inflow, which are the most effective at preventing the NTSI from developing, internal structures
form within the expanding slab interface, probably leading to fragmentation in the presence
of self-gravity or thermal instabilities. High Reynolds numbers result in local turbulence
 within the perturbed slab, which in turn triggers reconnection and dissipation of the 
excess magnetic flux. We find that when the magnetic field is initially aligned with the flow,
there exists a (weak) correlation between field strength and gas density. However, for transverse 
fields, this correlation essentially vanishes. In light of these results, our general conclusion is
that instabilities are unlikely to be erased unless the magnetic energy in clouds is much larger than
the turbulent energy. Finally, while our study is motivated by the scenario of molecular cloud formation
in colliding flows, our results span a larger range of applicability, from supernovae shells to colliding
stellar winds.
\end{abstract}
\keywords{instabilities --- MHD --- turbulence --- methods:numerical 
          --- ISM:clouds --- ISM: magnetic fields}

%
%
\section{Motivation}

Shocks and shells exist abundantly in the interstellar medium (ISM). 
Driven by supernovae, expanding HII-regions, gravitational 
flows or wholesale cloud collisions, they not only strongly influence the ISM
dynamics, but also affect its chemistry. 
However, structural analyses 
of the ISM show spectral indices closer to the Kolmogorov
value of incompressible turbulence (see \citealp{2004ARA&A..42..211E} for a review). 
Independently of the problem of how well these indices constrain the type of turbulence,
there are plenty of physical mechanisms to explain the 
closeness to the Kolmogorov value, ranging from the intrinsic nature of MHD turbulence 
(e.g. \citealp{1995ApJ...438..763G}, \citealp{2002PhRvL..89c1102B}, 
\citealp{2003MNRAS.345..325C}) to the 
conversion of compressible to solenoidal modes (\citealp{1994ApJ...436..728F}, 
\citealp{2004ARA&A..42..211E}). 

It is the latter mechanism that motivated this study. In the absence
of shear flows (oblique shocks) and thermal instabilities, the 
Non-linear Thin Shell Instability \citep{1994ApJ...428..186V} provides a natural
mechanism to convert compressible motions into solenoidal ones.
The NTSI is a rippling instability, relying on transverse momentum
transport due to bends in the collision interface of two opposing flows.
It is likely to arise in a wide range of environments, from colliding stellar winds,
supernova shells, colliding HI streams/clouds, to galaxy mergers.

The NTSI has been widely studied numerically (see \citealp{2006ApJ...648.1052H} for
a summary of the literature), mostly focusing on the effects of 
self-gravity and thermal instabilities. 
Our interest in the NTSI comes from the role it plays in the evolution
of molecular clouds in colliding HI flows 
\citep{2005ApJ...633L.113H,2006ApJ...643..245V,2006ApJ...648.1052H},
but the results have a wider applicability. With the exception
of \citet{1998ApJ...497..777K}, who included the magnetic pressure term in their
study of cloud collisions, work on the NTSI has so far neglected the 
effect of magnetic fields. However, fields could have a deciding influence on the 
evolution of a shock-bounded slab. Motivated by the numerical models of 
\citet{1995ApJ...441..702V} and \citet{1995ApJ...455..536P}, 
\citet{2001ApJ...562..852H} and \citet{2004ApJ...612..921B}
suggested that fields could, in fact, lead to a selection effect for
molecular cloud formation: clouds can only form if the fields are aligned
with the flows assembling the gas.

As a first step, we revisit the isothermal analysis of \citet{1994ApJ...428..186V} 
numerically, and study the evolution of the NTSI in a 
two-dimensional, magnetohydrodynamical environment. While the general
expectation (\S\ref{ss:physics}) is met in the laminar case, namely that 
magnetic fields can efficiently damp the NTSI, the geometry of the rippled interface induces  
non-ideal MHD effects, requiring a numerical method capable of handling
such effects in a stable and accurate way (\S\ref{ss:numerics}). 
However, we find that the exact amount of dampening crucially depends on the field 
orientation and strength (\S\ref{s:results}). This is 
especially true in the turbulent case, where turbulent reconnection inside the 
over-pressured slab leads to a pressure deficit, thus compressing the gas even further.
Finally, we show that the correlation between field strength and 
gas density is at best weak, even in the case of fields  
perpendicularly oriented with respect to the inflow, for which the field
would be expected to scale linearly with the density.

%
%
\section{Physics and Numerics\label{s:physnum}}

\subsection{Physics\label{ss:physics}}
The growth rate of the NTSI is mostly controlled by $k\eta$, the product
of the wave number of the slab perturbation $k$, and the amplitude of the 
slab's initial displacement $\eta$ (equivalently, the amplitude of the
collision interface's geometrical perturbation). 
The instability is driven by lateral transport of longitudinal momentum, 
i.e. if the inflow is parallel to the $x$ direction, and the slab is in the
$y$-$z$-plane, $x$-momentum is transported laterally in $y$ (and $z$),
collecting at the focal points of the perturbed slab. The efficiency of lateral
momentum transport is key to the development of the instability, since
it is the imbalance of ram pressure at the focal points that eventually propels
matter forward, driving the growth of the slab's perturbation. 
\citet{1994ApJ...428..186V} derived a growth rate of
\begin{equation} 
  \omega \approx c_sk (k\eta)^{1/2},\label{e:vishniac}
\end{equation}
with the sound speed $c_s$. \citet{1996NewA....1..235B}
found that at constant $\eta$ and for small $k$s, 
equation~(\ref{e:vishniac}) yields only a lower limit, while for large $k$s, the 
analytical growth rates agree well with the numerical results.
The reason for this seems to lie in the efficiency of deflecting the incoming
flow: for small $k$s, a small fraction of the incoming flow's momentum
is converted to lateral motions, while a large part compresses the 
slab (depending on the equation of state, this could lead to an increase in energy
losses). 

After the initial growth-phase (eq.~[\ref{e:vishniac}]), the NTSI reaches saturation
through two mechanisms: (i) expansion of the slab which stops the lateral momentum
transport by preventing the inflow from reaching the focal points, and (ii)
shear flow (Kelvin-Helmholtz) instabilities (KHI) in regions of the slab connecting
the focal points. The KHIs both generate inner structure and revive the slab's 
expansion. Note that strong cooling (not modeled here) can also suppress
the NTSI via early fragmentation \citep{2003NewA....8..295H}. 

Qualitatively, we expect magnetic fields to prevent the NTSI and 
subsequent KHI-modes from occurring. However, the detailed quantitative
extent of the damping should depend on the orientation
of the field with respect to the inflow. Indeed, fields aligned with the inflow
resist instabilities via the magnetic tension force, and therefore should be more efficient 
in suppressing the NTSI when $k\eta$ is small, even though the strong pairwise
field reversals arising from the opposed shear velocities along the slab (see \S\ref{ss:estimate})
could trigger reconnection. On the other hand, 
fields perpendicular to the inflow (but still in the 2D plane), primarily prevent 
instabilities from growing because of the magnetic pressure term in the Lorentz 
force, and to a lesser extent, because of
magnetic tension. For the sake of completeness, we mention that the third possible 
field configuration in 2D, i.e. the one in which the magnetic field is perpendicular 
to the flow plane, is irrelevant for this study since in that case the gas behaves 
as a system with an adiabatic exponent of $2$ for which the NTSI cannot be excited
\citep{1994ApJ...428..186V}.

\subsection{Numerics\label{ss:numerics}}
The magnetohydrodynamical scheme\footnote{We called the scheme PROTEUS, under 
which name we will refer to it subsequently. Proteus is a lesser 
god in Greek mythology, also known as "The Old Man of the Sea".
He lives in the sea off the coast of 
Egypt and can see things in the past, presence and future, but 
is very unwilling to share his knowledge. In order to evade questions, 
he has the ability to change his appearance. However, if you manage
to catch and hold him, he will assume his true shape and answer 
your questions.} is based on a conservative gas-kinetic flux-splitting 
method, introduced by \citet{1999JCoPh.153..334X} and 
\citet{2000JCoPh.165...69T} and derived from the 1st-order BGK 
\citep{1954PhRv...94..511B} model. Representing
the velocity distributions as Maxwellians in each cell, fluxes
across cell walls are derived from the differences in the velocity moments of
Maxwellian distributions reconstructed at the cell walls.
The reconstruction is second order in space using MUSCL limiters,
and it allows a fast and consistent way to implement 
viscosity and Ohmic resistivity in the form of dissipative
fluxes \citep{2004ApJ...603..165H} at close to zero extra computing cost, while preserving the
time order of PROTEUS since the dissipative terms are not simply added as source terms 
but are part of the flux computation. This allows us to control dissipation
in a physical manner, without having to rely on numerical dissipation
to terminate the turbulent cascade at grid scale. Total energy is conserved
at machine-accuracy level for an adiabatic equation of state. In the isothermal
version which we are using in \S\ref{s:results}, the total energy equation is not evolved. PROTEUS uses a 
2nd order TVD Runge-Kutta time stepping \citep{1988JCP..77..439S} for the MHD equations
to achieve 2nd order temporal accuracy. Fluxes are updated
in time-unsplit fashion, i.e. flux updates for spatial directions
are computed using the initial conditions of the current time step. 
In order to keep $\nabla\cdot\mbfB=0$, PROTEUS employs 
a Hodge projection \citep{1994JSSC..15..263Z,1998ApJS..116..133B}. The code is fully 
message passing interface (MPI) parallelized. 

With PROTEUS, one may switch between the
MHD-solver previously described and a purely hydrodynamical solver based on
the 2nd-order BGK model. The latter implementation has been 
introduced and extensively discussed by \citet{1993JCoPh.109...53P}, \citet{1999A&AS..139..199S}, 
\citet{2006ApJ...648.1052H} and \citet{slyz2006submit}, and we therefore refer the reader interested 
in implementation details to these papers.

One-dimensional shock tests
and the Orszag-Tang vortex have already been discussed by \citet{2000JCoPh.165...69T}, 
hence in what follows, we focus on three other MHD test cases.
The two first ones, i.e. the propagation of a linear
Alfv\'{e}n wave under resistive damping (\S\ref{ss:linalfven})
and the current sheet evolution (\S\ref{ss:currentsheet})
are both meant to test the resistive fluxes, while the third one, i.e.
the advection of a field loop (\S\ref{ss:fieldloop}) is a geometry test.
A detailed study of the magnetized Kelvin-Helmholtz
Instability is under way \citep{palotti2006submit}.

\subsubsection{Propagation of a linear Alfv\'{e}n Wave\label{ss:linalfven}}
\begin{figure}
  \includegraphics[width=\columnwidth]{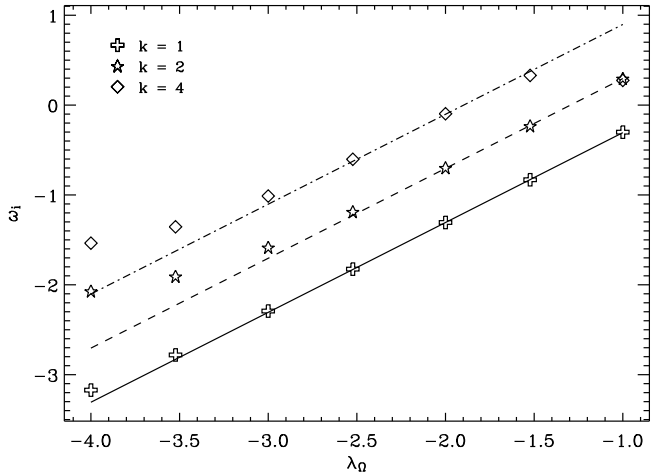}
  \caption{\label{f:alfven}Damping rate (eq.~[\ref{e:alfven}]) of a
           linear Alfv\'{e}n wave against Ohmic resistivity for 
           $\kappa=1,2,4$. The resolution is $N=64$. Errors of the
           measured damping rates are smaller than the symbol sizes.
           Lines denote the analytical solution.}
\end{figure}

This one-dimensional test checks the resistive flux implementation
as well as the accuracy of the overall scheme. 
A linear Alfv\'{e}n wave under weak Ohmic dissipation is damped
at a rate of
\begin{equation}
  \omega_i = \f{1}{2}\lomega k^2,\label{e:alfven}
\end{equation}
where $\lomega$ is the Ohmic resistivity, and $k=2\pi\kappa/L$ is the 
wave number of the Alfv\'{e}n wave, with $\kappa\in\mathbb{N}$. 
The strongly damped case, where the decay dominates the time evolution,
is physically uninteresting for our application, since the Ohmic 
resistivity is mainly used to control numerical dissipation.
Figure~\ref{f:alfven} shows the damping rate against Ohmic resistivity 
$\lomega$ for $\kappa=1,2,4$ at a grid resolution of $N=64$. 

From Figure~\ref{f:alfven}, it is clear that, as one diminishes the value of  
 $\lomega$, there comes a point when the numerical resistivity of the scheme
becomes comparable to the physical one, causing the measured damping rate to 
flatten out and depart from the analytical solution. For $\kappa=4$ and 
$\lomega=0.1$, the wave decays too quickly to allow a reliable measurement, 
and the system enters the strongly damped branch of the dispersion relation. 
However, we emphasize that, even for this high value of $\kappa$ in light of 
the modest resolution we used, the resistivity range available to PROTEUS 
spans three orders of magnitude.

\subsubsection{Current Sheet\label{ss:currentsheet}}
This test is taken from \citet{1995CoPhyCom.89..127H} and the ATHENA test 
suite\footnote{{\tt http://www.astro.princeton.edu/\\
                   $\sim$jstone/tests/field-loop/Field-loop.html}}.
A square domain of extent $-0.5\leq x,y\leq 0.5$ and of constant density $n_0=1$ 
and pressure $p_0$ is permeated by a magnetic field along the $y$ direction such that
$B_y(|x|<0.25)=\sqrt{4\pi}$, and $B_y=-\sqrt{4\pi}$ elsewhere. This results in two 
magnetic null lines, which then are perturbed by velocities $v_x=A\sin(2\pi y)$.
The goal is to find the pressure $p_0$ and velocity amplitude $A$ for which the code
crashes. The main problem is -- especially in conservative schemes -- that the
resistive decay of the field leads to strong localized heating that in turn
generates strong magnetosonic waves. Thus, the smaller $p_0$ and/or the larger
$A$, the harder the test. We chose $p_0=0.1$ and $A=0.3$ fairly close to 
the ``standard'' values quoted on the ATHENA web site$^6$, $p_0=0.05$ and $A=0.1$. 
Here, we use an adiabatic exponent of $\gamma=5/3$ and employ the 
conservative formulation of the scheme.

\begin{figure}
  \includegraphics[width=\columnwidth]{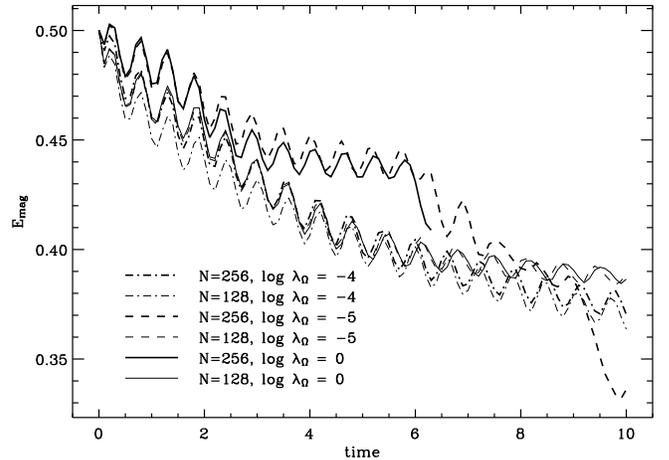}
  \caption{\label{f:emagtime}Total magnetic energy against time for the current sheet 
           test. A finite resistivity $\lomega$ helps stabilize the code.}
\end{figure}

The test turns out to be very hard for PROTEUS, because of its low
numerical diffusivity. Figure~\ref{f:emagtime} summarizes the test results
in the form of the total magnetic energy against time. The energy evolution
splits into two branches: one corresponding to the lower-resolution models at
all resistivities, and the other representing the higher-resolution models
at $\lomega=0$ and $\lomega=10^{-5}$. At $\lomega=10^{-4}$, the field decay has
converged: both resolutions give the same curve. At $N=256^2$ and $\lomega=0$,
the code crashes around $t=6$. All other models run up to $t=10$ and further.

The result confirms the discussion by \citet{2000JCoPh.165...69T}, namely that while 
the conservative gas-kinetic flux splitting method in the BGK-formalism performs well 
for high-$\beta$ plasmas (with $\beta$ defined as the ratio of thermal over magnetic pressure), 
it might not be the method of choice for low-$\beta$ plasmas, i.e.
magnetically dominated systems. Note, however, that this is mainly a consequence of the 
scheme's conservative formulation.
We experimented with a non-conservative version (i.e. just evolving the internal energy
instead of the total energy), which was stable for lower $p_0$ and higher $A$, albeit at the cost
of a less accurate total energy evolution.

\subsubsection{Advection of a Field Loop\label{ss:fieldloop}}

\begin{figure}
  \includegraphics[width=\columnwidth]{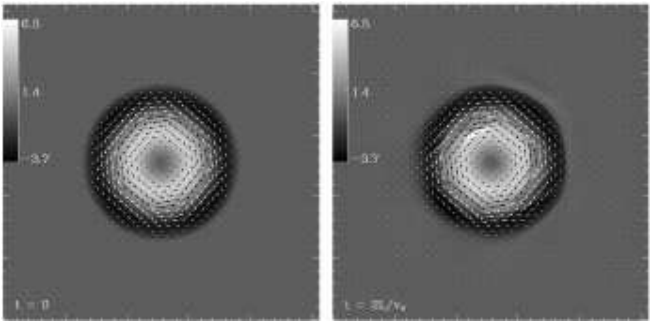}
  \caption{\label{f:fieldloop}Current density for field loop advection test 
          (see \S\ref{ss:fieldloop}).
  {\em Left:} Initial condition. {\em Right:} After two horizontal crossings. The grid resolution
   is $256^2$.}
\end{figure}

A cylindrical current distribution (i.e. a field loop) is advected diagonally across
the simulation domain. We follow the implementation presented by \citet{2005JCoPh.205..509G}
and the ATHENA test suite$^6$,
based on an earlier version by \citet{1996JCoPh.128...82T}. Density and pressure are both initially
uniform at $n_0=1$ and $p=1$, and the fluid is described as an ideal gas with an adiabatic exponent 
of $\gamma=5/3$. The square
grid ranges from $-0.5\leq x \leq 0.5$, and the loop is advected at an angle of $30$ degrees
with respect to the $x$-axis. Thus, two round trips in $x$ correspond to one crossing in $y$. 
The amplitude of the field loop is set to values $10^{-3}$, $10^{-2}$ and $10^{-1}$, with an
initial radius of $R_0=0.3$. Figure~\ref{f:fieldloop} shows the initial current distribution
with the magnetic field vectors over-plotted ({\em left}), and the final current
distribution after two time-units measured in horizontal crossing times.
The overall shape is preserved, although some artifacts are visible at the upper
rim of the loop. These results concerning the shape are qualitatively similar 
to those posted on the above mentioned website.
Note that we show the current density, since the artifacts do not show up in the
magnetic energy. This test uses $\lomega \equiv 0$. 

\begin{figure}
  \includegraphics[width=\columnwidth]{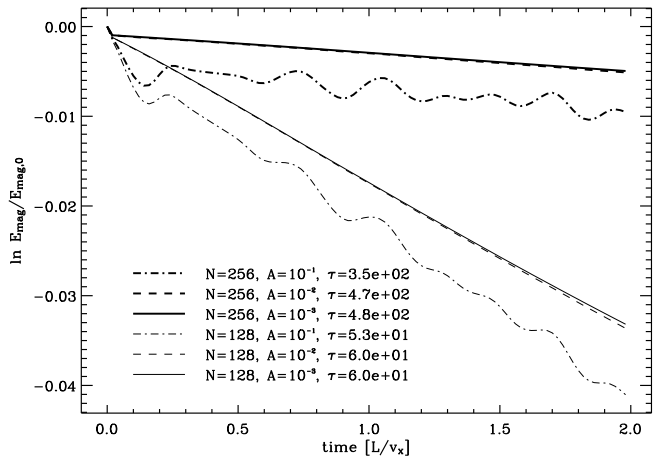}
  \caption{\label{f:decayloop}Normalized magnetic energy against time 
  (in units of horizontal crossing time), for two resolutions
  and three field strengths. The highest amplitude leads to waves when perturbed
  by advection. At a resolution of $128^2$, the magnetic energy has decayed by $3.3$\% after
  two crossing times. The fit decay times $\tau$ are indicated for each model.}
\end{figure}

Figure~\ref{f:decayloop} presents a quantitative diagnostic of the behavior of the code as it tracks
the magnetic energy decay against the simulation time,
in units of horizontal crossing times. The initial energy is normalized to one.
Line styles denote different amplitudes of the field strength. The solid lines correspond
to the case given by \citet{2005JCoPh.205..509G}, the dashed and dash-dotted lines denote cases with larger
amplitudes. At an amplitude of $A=10^{-3}$ and a resolution of $128^2$, the magnetic
energy decays by $3.3$\% over two horizontal crossing times. The ATHENA website$^6$ quotes a decay
of $3.5$\% with a $256x148$ grid, using a Roe solver
and 3rd order reconstruction.

In summary, these numerical test cases demonstrate that PROTEUS models dissipative MHD effects
accurately, due to a low intrinsic numerical diffusivity that compares well with that of higher-order
Godunov methods. Furthermore, it can advect geometrically complex magnetic field patterns properly, and 
is well suited in its energy conserving form to model MHD flows with $\beta > 1$.

\subsection{Initial Conditions\label{ss:initcond}}

To remain as close as possible to \citet{1994ApJ...428..186V}, we will use the isothermal
version of PROTEUS in the following. The initial conditions are similar to those
discussed in \citet{2005ApJ...633L.113H,2006ApJ...648.1052H}. 
Two uniform, identical flows in 
the $x$-$y$ computational plane initially collide head-on at a sinusoidal 
interface with given wave number $k_y$ and amplitude $\eta$. The field is either
aligned or perpendicular to the inflow, but in both cases in the $x$-$y$ plane. 
For the standard runs, we used a rectangular grid with an extent of $88$~pc
in $x$ and $44$~pc in $y$. Field strength as well as viscosity and resistivity
are varied.  The grid resolution varies between $N_x\times N_y=256\times 128$ and
$2048\times 1024$ by factors of $2$ in linear resolution. 
The isothermal sound speed is $c_S = 5.3$ km s$^{-1}$, the
Mach number of the instreaming gas is ${\cal M} = 4$, and the inflow density
is set to $n_0 = 1$ cm$^{-3}$. Thus, in the code unit system, the Alfv\'{e}n speed
in the inflow region is given by $c_A=B$, the magnetic field strength.  

%
%
\section{Results\label{s:results}}

We give a rough estimate for the field strength required to prevent the 
excitation of the NTSI in \S\ref{ss:estimate}.
The morphology of the instability naturally depends strongly on the field
orientation. We present some examples in \S\ref{ss:morphology}. Because
of the strong shear flows, the explicit control of dissipation is
crucial in reaching numerical convergence. This can be further quantified by
monitoring the growth rates (\S\ref{ss:growthrates}). Finally, we show that the geometry of the 
flow and magnetic field strongly influence the field-density relation
(\S\ref{ss:fielddensity}).

\subsection{Estimate of Threshold Field Strength\label{ss:estimate}}

\begin{figure}
  \includegraphics[width=\columnwidth]{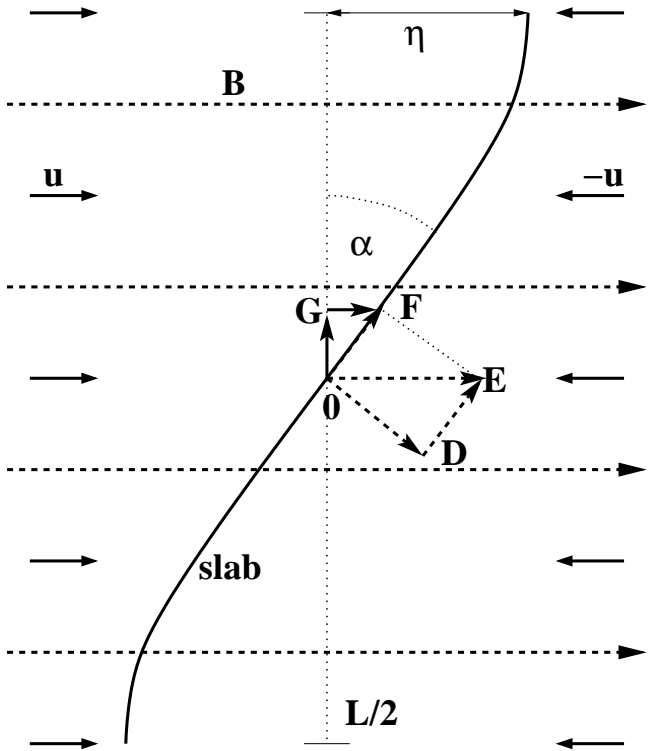}
  \caption{\label{f:sketch}Sketch of force geometry for estimating
           the threshold field strength. A thick solid line marks the 
	   interface of the colliding flows (the ``slab''). The magnetic 
           field is represented by horizontal 
           long short-dashed arrows, and inflow velocities by short solid ones.
           Other (force) vectors are defined in the text.}
\end{figure}

A very rough estimate of the threshold field strength preventing the
excitation of the NTSI can be derived by simple pressure considerations.
Figure~\ref{f:sketch} gives a sketch of the simplified situation.
Only one half of the slab in the vertical direction is shown. The slab
is displaced by $\eta$ in the horizontal direction around the 
(dotted) center line. The angle
between the slab and the symmetry line measured at point $0$ is
given by $\alpha$, with $\tan\alpha\approx 2\eta k/\pi$. 
Gas is streaming in horizontally from the left and the right
with velocity $\mbfu$, and the magnetic field $\mbfB$ is aligned with
the inflow, pointing to the right. Incoming flow with positive velocities
exerts a pressure at point $0$, which can be split into a normal
component $\overline{0D}$ and a tangential component 
$\overline{0F}=\overline{0E}\,\sin\alpha$.
The tangential component corresponds to the ram pressure exerted
by material deflected by the slab and sliding along $\overline{0F}$. 
To a zeroth order approximation, the component of this ram pressure 
perpendicular to $\mbfB$ is available for bending the field lines, thus
\begin{equation}
  \rho\,u^2\,\sin\alpha\,\cos\alpha \approx \mbfB^2/2.\label{e:crudeest}
\end{equation}

The maximum is reached for $\alpha = \pi/4$, in which case
$\rho u^2 \approx \mbfB^2$. For an inflow velocity of $|u| = 4 c_s$, 
we thus require a field strength corresponding to 
an Alfv\'{e}n speed of $c_A\approx 4 c_s$ to suppress the NTSI. 

\subsection{Morphology\label{ss:morphology}}

We begin with our standard runs, in order to compare to Vishniac's
analysis. These are the ``laminar'' cases (\S\ref{sss:laminar}), i.e.
cases that do not develop turbulent substructure. The turbulent case
and the relevance of fixed physical dissipation scales are discussed
in \S\ref{sss:turbulent}.

\subsubsection{Laminar Case\label{sss:laminar}}
\begin{figure}
  \includegraphics[width=\columnwidth]{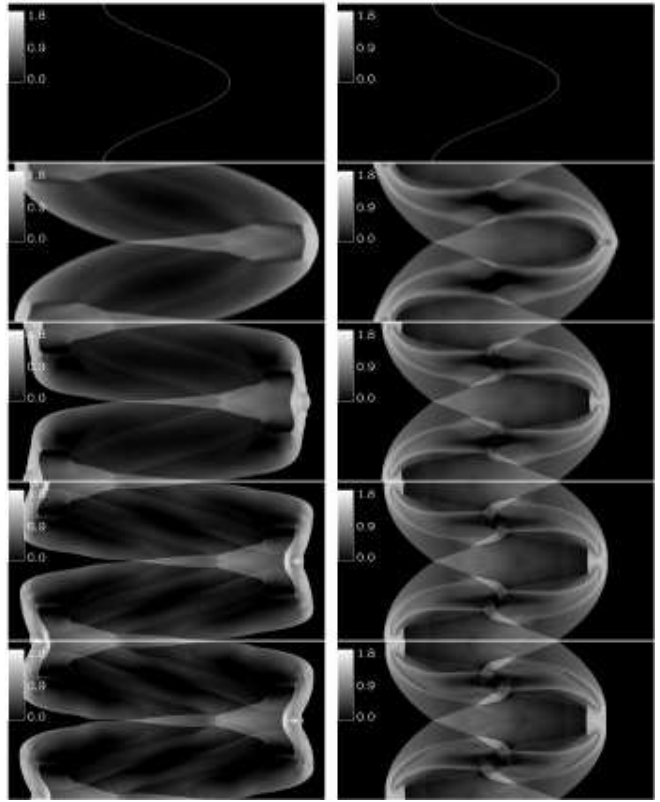}
  \caption{\label{f:denshdmhd}{\em Left column:} Logarithmic density maps of 
          hydrodynamical models at $t=3.8$~Myr corresponding to the end of simulation, 
          resolution increasing
          from top to bottom ($N_x=256$ to $2048$ by factors of $2$). The two 
          highest resolutions have 
          converged even in detail. {\em Right column:} Same as left, but for the magnetic 
          models, where the field is
          aligned with inflow and $c_A/c_s=1.0$. Again, the two highest resolutions have
          converged.}
\end{figure}

The left column of Figure~\ref{f:denshdmhd} displays a model sequence in 
resolution for the hydro runs, corresponding to Vishniac's analysis. The top
panel shows the initial condition (strictly, just after $t=0$), 
The increasing overall amplitude of the slab signifies that the NTSI is clearly at work. 
Most of the gas is collected at the focal
points, and by the end of the simulations, the system is close to saturation.
The lowest resolution run differs from all others in that it is the only one for which
numerical convergence has not been
achieved. The two highest resolution runs ($N_x=1024,2048$, lower two panels), 
on the other hand have converged even in detail.
Note that the viscosity and resistivity provide fixed 
physical dissipative scales, independent of the resolution. Thus, the model at
$N_x=256$ is not resolved with respect to these dissipative scales, while
the high-resolution models are.

\begin{figure*}
  \includegraphics[width=\textwidth]{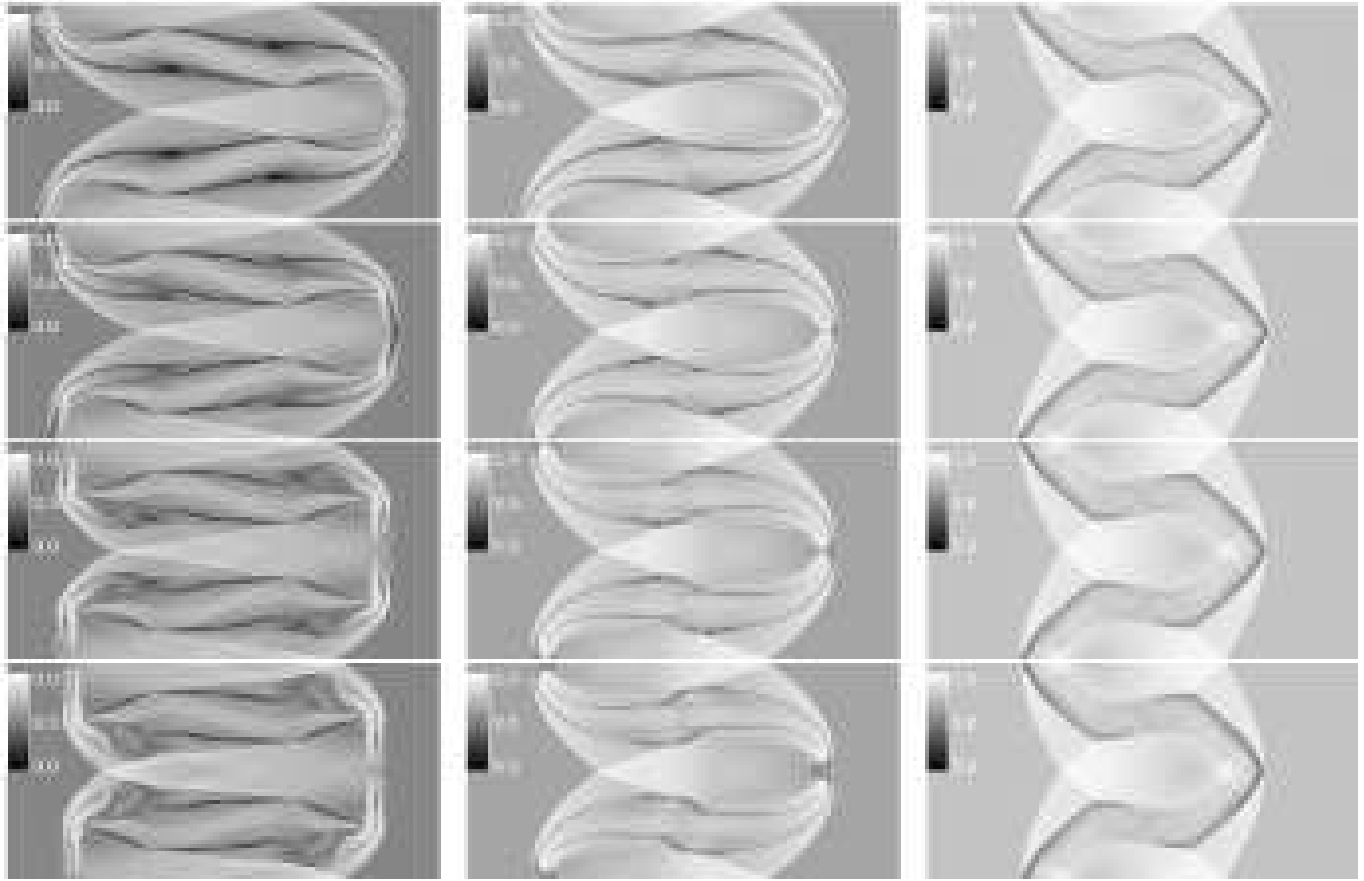}
  \caption{\label{f:emag}Logarithm of magnetic energy at $t=3.8$~Myr. {\em Left:} Four resolutions 
          increasing from top to bottom ($N_x=256$ to $2048$ by factors of $2$) at $c_A/c_s=0.5$. 
          {\em Center:} For $c_A/c_s=1.0$. {\em Right:} For $c_A/c_s=2.0$.}
\end{figure*}

This is also true for the magnetic runs (right column of Figure~\ref{f:denshdmhd}) where 
the field has slowed down the growth of the NTSI. The resulting slab is also more structured than in the
pure hydro case. High-density regions, especially thin filaments, coincide with regions
of field reversals (loss of magnetic pressure support). To see this, one can compare 
the gas density to the magnetic energy (Figure~\ref{f:emag}) maps. The center column corresponds to the 
right column of Figure~\ref{f:denshdmhd}. Not only do the field reversals lead to 
dense structures, but magnetic waves arise. The left column of 
Figure~\ref{f:emag} shows the same resolution sequence at half the field strength,
i.e. $c_A/c_s = 0.5$, while the right column stands for $c_A/c_s = 2.0$. 
Higher field strengths not only reduce the growth rate of the NTSI, but also suppress 
the internal turbulent structure visible for $c_A/c_s=0.5$. Numerical convergence
is more easily achieved with higher field strength, which is another indicator that 
turbulence plays a minor role, i.e. we remain safely entrenched within the laminar regime.
This is slightly different in the weak-field case where the two highest resolution runs
have only mildly converged. Here, the field starts to get too weak to prevent the
excitation of KHI modes. 

\begin{figure}
  \includegraphics[width=\columnwidth]{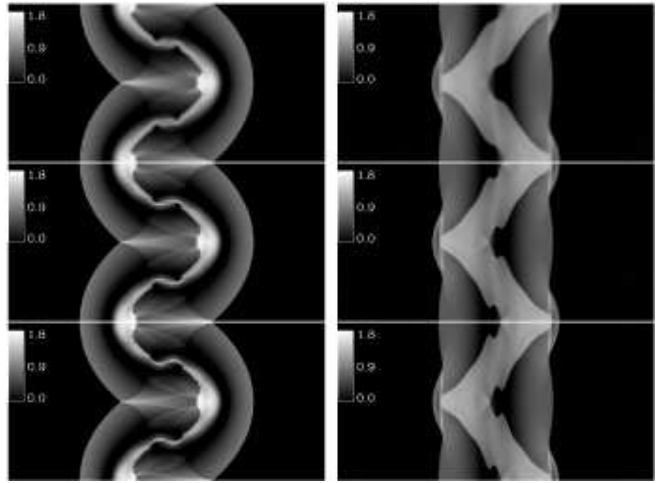}
  \caption{\label{f:densy}Logarithmic density maps for models with the field oriented 
          transversally, i.e. perpendicular to the
          inflow, at $t=3.8$~Myr. {\em Left:} Three resolutions increasing from top to bottom 
          ($N_x=256$ to $1024$ by factors of $2$) at $c_A/c_s=0.5$. {\em Right:} For $c_A/c_s=1.0$.
          The geometric pattern at the boundaries is an artifact caused by magnetosonic 
          modes reaching the inflow boundaries.}
\end{figure}

\begin{figure}
  \includegraphics[width=\columnwidth]{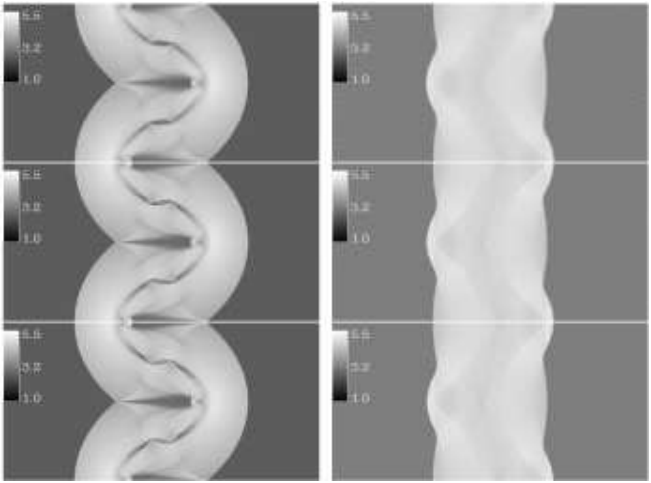}
  \caption{\label{f:emagy}Logarithmic magnetic energy maps of models with 
          the field oriented transversally, i.e. perpendicular to the
          inflow, at $t=3.8$~Myr. {\em Left:} Three resolutions increasing from top to bottom
          ($N_x=256$ to $1024$ by factors of $2$) at $c_A/c_s=0.5$. {\em Right:} For $c_A/c_s=1.0$.}
\end{figure}

A variation on the theme is shown in Figures~\ref{f:densy} and \ref{f:emagy} 
where the field is oriented
vertically this time around. In a 1D geometry, the magnetic pressure would prevent the 
gas from efficiently accumulating to form high-density regions (i.e. clouds, see \citealp{2004ApJ...612..921B}). 
The system would behave as if the gas had an adiabatic exponent of $2$. This is the situation traced
out by the models shown in the right column of Figures~\ref{f:densy} and \ref{f:emagy}:
because of the stiffened equation of state, high (flux-)density regions expand faster. However,
high density is found at the focal points, hence the initial perturbation of the slab
is smoothed out, and the slab ends up with plane-parallel shock fronts (due to the fields,
there are some waves inside the slab, though). Reducing the field strength (left column) by
a factor of $2$ however changes the situation drastically. Although the NTSI is only weak,
high-density filaments start to form in the thick slab, again at the locations
of field reversals. Thus, a transverse field is much less of an inhibiting factor for
substructure or high-density region generation than what would be expected from a 
1D-argument. For $c_A/c_s = 1.0$, the fast magnetosonic modes have already reached the boundaries,
causing the geometric patterns visible in the density plot.

\subsubsection{Turbulent Case\label{sss:turbulent}}

In the previous section, we discussed models with a fixed physical dissipative scale.
The purpose of this section is to demonstrate that without a fixed dissipation scale, 
numerical convergence cannot be reached. In other words, for large Reynolds numbers, 
the system can evolve qualitatively differently.

\begin{figure}
  \includegraphics[width=\columnwidth]{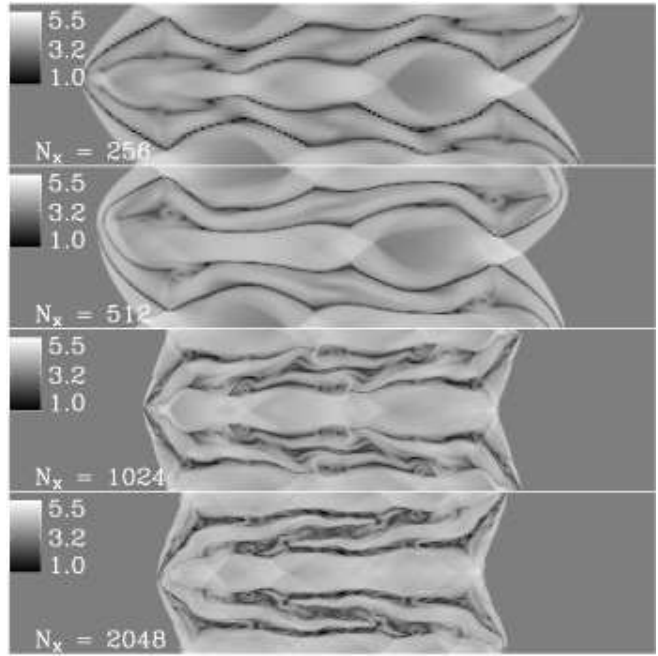}
  \caption{\label{f:turb}Logarithmic magnetic energy maps for models with 
           viscosity $\nu\equiv 0$ and resistivity $\lambda_\Omega=0$, i.e.
           the dissipation scale is given by the grid resolution.
           Resolution increases from top to bottom
           ($256^2$ to $2048^2$ by factors of $2$). Since the period in $y$ is
           repeated four times, we show only one quarter of the domain. $t=3.8$~Myr.}
\end{figure}

Figure~\ref{f:turb} shows a sequence in resolution of models with zero physical resistivity and
viscosity, i.e. models for which the numerical dissipation at resolution scale will set the 
Reynolds number. Thus, higher resolution will lead to larger Reynolds numbers. 
For resolution reasons we chose the wave number of the interface perturbation to be
$k=1$ in \S\ref{sss:laminar}. Since the condition for fast growth of the instability
is given by $k\eta\approx 1$, this required a larger initial amplitude perturbation and an elongated box. 
Here, we are interested in the (later) turbulent evolution of the slab, thus we start with $k=4$, 
which allows us to reduce the initial amplitude of the perturbation by the same factor $4$ 
and therefore considerably extends the spatial range in which the slab can develop.

At our lowest resolution ($N_x=256$), we essentially
get a laminar behavior: the pairwise field reversal regions are stretched along the width
of the slab but persist to the end of the simulation (top). At $N_x=1024$,
the main magnetic null regions are accompanied by secondary regions as 
a result of additional shear flows. The slab is thinner. Increasing the resolution
further introduces more and more substructure in the slab, especially at 
the ``heads'' of the slab's perturbations: here, field reversals seem to accumulate,
leading to additional field dissipation. Thus, with higher Reynolds numbers, 
the slab gets more turbulent, and reconnection proceeds not only in the two
main magnetic null regions, but all throughout the slab in small regions.
Consequently, turbulent field structures inside the slab are dissipated faster, leading
to a deficit in magnetic pressure inside the slab, and thus to a thinner slab.

While this effect is certainly interesting to note, the situation might be less
extreme in a truly three-dimensional system: a third field component without
magnetic null could give rise to sufficient magnetic pressure to prevent
reconnection (see also \citealp{2003ApJ...590..291H}). In this case, small-scale
fields entangled by the turbulence in the slab could actually lead to 
additional pressure.

\begin{figure}
  \includegraphics[width=\columnwidth]{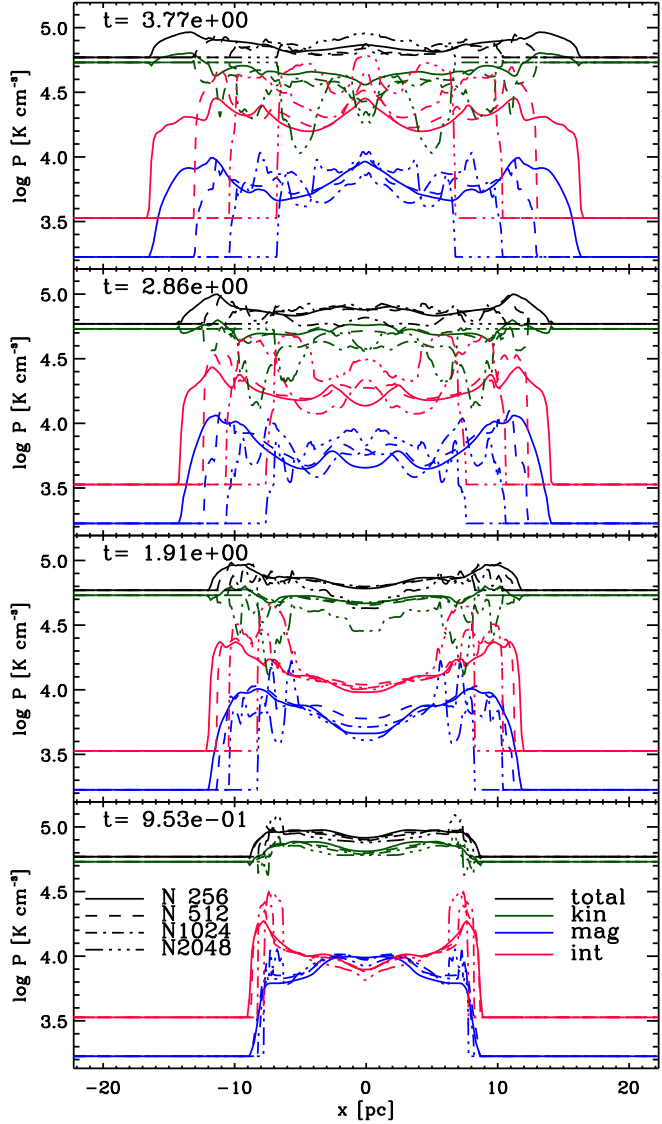}
  \caption{\label{f:pressprof}Transversally averaged pressure profiles for 
           four times (in Myr), plotted in ascending time order
           from bottom to top, against $x$-axis coordinate. 
           Shown are pressure profiles for magnetic models with $\nu\equiv 0$ and
           $\lomega\equiv 0$ (see text).
           Gas is streaming in from the left
           and from the right. See bottom panel for color code and line styles.
           Shown are the total, the kinetic, the magnetic and the internal pressures.}
\end{figure}

\begin{figure}
  \includegraphics[width=\columnwidth]{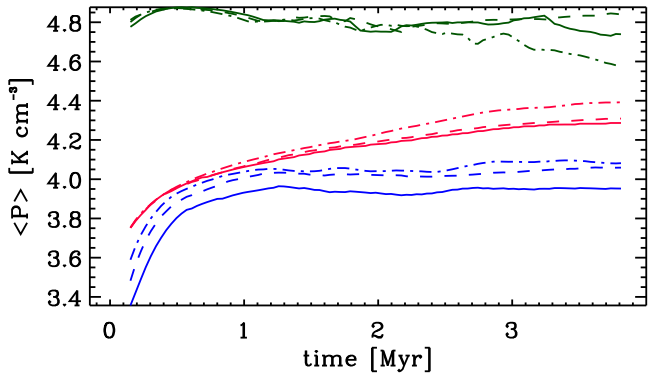}
  \caption{\label{f:pressequi}Average pressure in slab against time. Color coding
           and line style are the same as in Figure~\ref{f:pressprof}.}
\end{figure}

The pressure profiles (Fig.~\ref{f:pressprof}) actually tell us a slightly more 
complicated story than that of simple magnetic energy dissipation. Pressures were
averaged transversally (i.e. along the $y$-axis) and plotted against $x$, the inflow
direction.
The magnetic pressure profile stays pretty much at a constant
level, independent of resolution and time. What changes is the kinetic pressure,
which drops below its inflow value (all panels but bottom). This effect gets
stronger with increasing resolution. Thus, the magnetic field only acts as a 
dissipation channel for the kinetic energy. One could even interpret the 
kinetic pressure drop and simultaneous magnetic and internal (red lines)
pressure rise as an attempt of the system to achieve equipartition (Fig.~\ref{f:pressequi}). 

\begin{figure}
  \includegraphics[width=\columnwidth]{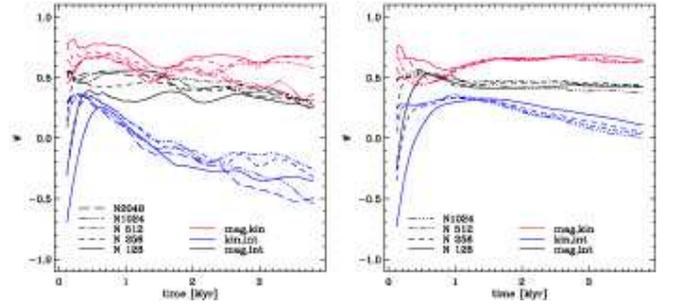}
  \caption{\label{f:presscorrel}Correlation of pressures within slab against time. 
           Line colors
           denote correlations, e.g. red stands for ${\cal C}(P_{mag},P_{kin})$. 
           Line styles denote resolution. {\em Left:} Models without fixed
           physical dissipation scales. {\em Right:} Viscous and resistive scales
           have been set.}
\end{figure}

This is not to say that the various pressure components are in equilibrium, 
as the left panel of Figure~\ref{f:presscorrel} easily demonstrates. 
This panel shows the correlation coefficient ${\cal C}$ for the three pairs of pressures,
$P_{mag}$, $P_{int}$, and $P_{kin}$, for the case without (left) and with (right)
a physical dissipation scale. Balance between the pressure components would show
up as an anti-correlation, whereas a correlation can be interpreted as pressures being in phase 
(and thus driving waves). Decorrelated pressures indicate a mixture.
Kinetic and magnetic pressure decorrelate at higher resolution,
pointing to strong reconnection events. Internal and magnetic pressure are only slightly
correlated (see below), while kinetic and internal pressure anti-correlate at late
times, because high-density regions show more inertia. The models with a fixed
physical dissipation scale do not show strong resolution effects.
Note, however, that the ``dissipation-less'' models are farther in the dynamical
evolution than the ``controlled'' models, because of the different initial
conditions ($k_y=4$ against $k_y=1$).

These results demonstrate that only a fixed dissipative scale can guarantee full convergence 
of the models with resolution. Relying on numerical diffusion leads to flows with increasing 
Reynolds numbers as the resolution increases and results that depend qualitatively and quantitatively 
on resolution. 

\subsection{Growth Rates\label{ss:growthrates}}

\begin{figure}
  \includegraphics[width=\columnwidth]{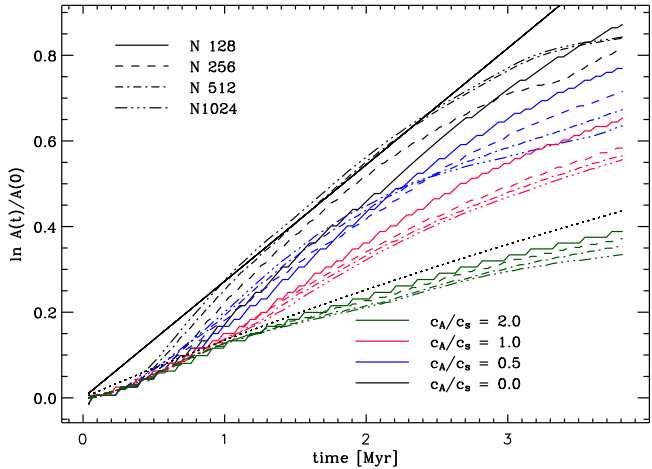}
  \caption{\label{f:growthrates_fld}Slab amplitude against time for four models
          at zero field strength $c_A/c_s = 0$ to highest field strength $c_A/c_s=2$.
          Line styles stand for resolution. The straight solid black line denotes
          the analytical solution (eq.[\ref{e:vishniac}], \citealp{1994ApJ...428..186V}). 
          The dotted line denotes the 
          expansion of a shock-bounded unperturbed slab (eq.~[\ref{e:unpertslab}]).}
\end{figure}

Figure~\ref{f:growthrates_fld} summarizes the growth of the slab's amplitude with
time. The hydrodynamical growth rates are consistent with the analytical predictions
(eq.~[\ref{e:vishniac}], solid straight black line). Saturation sets in when the 
focal points are shut off from the inflow (see also left column of Fig.~\ref{f:denshdmhd}).
The two highest resolution runs have converged also in terms of growth rates (we established 
detailed convergence in \S\ref{sss:laminar}). Lower resolutions lead to slightly smaller amplitudes
initially. The lowest resolution model is already resolved high enough to reproduce the laminar
result, but the specified physical viscosity is too small to guarantee convergence
with regards to turbulent substructure in the slab. Thus, when increasing to the next
higher resolution, the unresolved turbulence leads to less converged growth rates. 
Note that the growth rate as given
in equation~(\ref{e:vishniac}) does not include the effect of turbulence generation 
within the slab, although the possible effects of turbulence are discussed by \citet{1994ApJ...428..186V}. 

The dotted line in Figure~\ref{f:growthrates_fld} and all subsequent
figures of the same type denotes the growth of the amplitude of an unperturbed shock-bounded
slab of width $\Delta(t)$, 
\begin{equation}
  \Delta(t) = 2\,c_s\,t\left({{\cal M}/2+(1+({\cal M}/2)^2)^{1/2}}\right)^{-1}\label{e:unpertslab}.
\end{equation}
Thus, all models except for those with field strength $c_A/c_s=2$ show a faster growth of the slab
amplitude than just the shock-bounded expansion. Obviously, equation~\ref{e:crudeest} 
only gives a rough estimate of the field strength required, and in fact underestimates
the effect of the field. Our crude approximation treats the slab as a solid wall, 
allowing to split the pressure exerted by the inflow on the wall along the normal
and tangential directions without any losses. This is most likely unrealistic, i.e. the
tangential component will generally be smaller, thus requiring a smaller field
to balance it.

With increasing field strength, the ordering of the curves with respect to resolution
is inverted: now, the lowest-resolution runs show the fastest growth. Nevertheless
we get convergence for the two highest resolution simulations. This inversion is a consequence
of the slight increase in the Reynolds number with respect to the runs at lower resolution.
As discussed above, higher Reynolds numbers lead to more turbulence and more
field reversals. Thus, the field is dissipated faster, leading to a pressure deficit
in the slab.

\begin{figure}
  \includegraphics[width=\columnwidth]{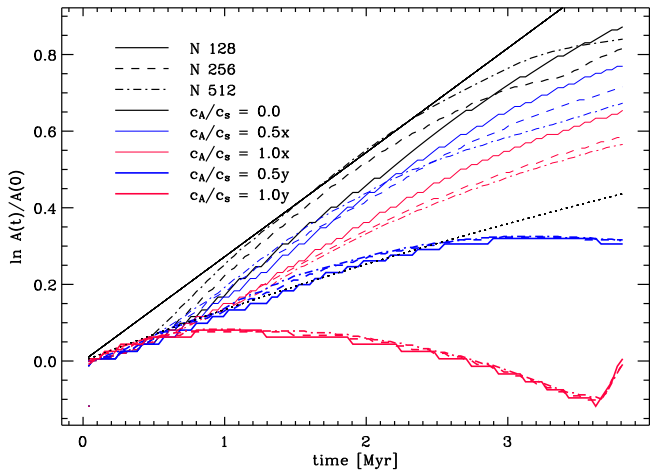}
  \caption{\label{f:growthrates_xy}Slab amplitude against time for models with transverse
          fields. 
          Thin lines denote the corresponding models with fields oriented along the inflow
          for comparison. Again, the straight solid black line denotes
          the analytical solution (eq.~[\ref{e:vishniac}], \citealp{1994ApJ...428..186V}). 
          The dotted line denotes the 
          expansion of a shock-bounded unperturbed slab (eq.~[\ref{e:unpertslab}]).}
\end{figure}

A vertical field leads to a strongly reduced growth rate or suppresses the instability
completely, depending on its strength (Figure~\ref{f:growthrates_xy}). The NTSI
arises, however, for weaker field strengths, and the bending mode in the slab
persists, thus allowing material to be funneled towards the focal points. 
In the presence of self-gravity or a strong thermal instability (e.g. that
provided by atomic line cooling), these density enhancements could then 
fragment and generate further substructure, despite the initially
unfavorable field orientation.

\begin{figure}
  \includegraphics[width=\columnwidth]{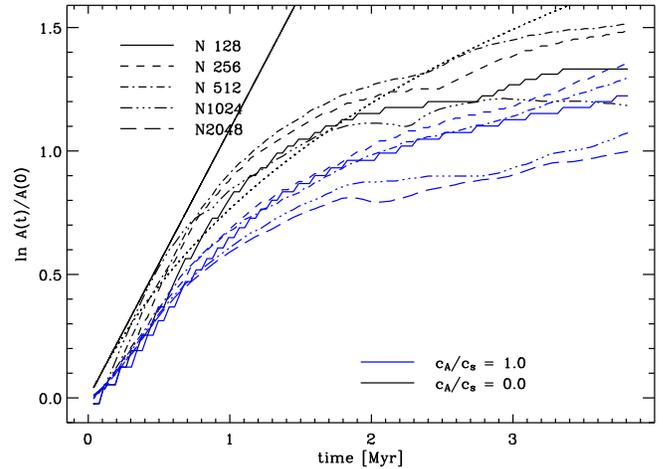}
  \caption{\label{f:growthrates_k4}Slab amplitude against time for models with initial perturbation
          wave number $k=4$ and without fixed resistive scale. Higher resolution leads to 
          saturation of the growth rate at lower amplitudes. The straight solid black line denotes
          the analytical solution (eq.~[\ref{e:vishniac}], \citealp{1994ApJ...428..186V}). 
          The dotted line denotes the 
          expansion of a shock-bounded unperturbed slab (eq.~[\ref{e:unpertslab}]).}
\end{figure}

The amplitude growth corresponding to Figure~\ref{f:turb} is shown in 
Figure~\ref{f:growthrates_k4}. With higher resolution, saturation sets in earlier in the 
magnetic runs, and the 
amplitudes even decrease with time, mirroring the loss of pressure inside the slab
due to (resistive) dissipation.
Note that the slab amplitude now grows faster than that of an unperturbed
shock-bounded slab (see dotted line in Fig.~\ref{f:growthrates_k4}) only at times $t\lesssim 1$~Myr. As 
Figure~\ref{f:turb} already had made us suspect, saturation sets in earlier for models
with higher $k$. 

\subsection{Field-Density Relation\label{ss:fielddensity}}

The field-density relation $B(n)$ is often used as an observational 
measure for the dynamical importance of magnetic fields in the interstellar
medium. It describes the mass-loading of field lines, and is related
to the mass-to-flux ratio (e.g. \citealp{1976ApJ...210..326M};
\citealp{1987ARA&A..25...23S})
quantifying the importance of magnetic fields in a gravitationally 
dominated cloud. Microscopically, the mass loading of field lines
can be changed in two ways: by Ohmic diffusion and by ion-neutral (or
ambipolar) drift. However, the two parameters controlling the degree
to  which the interstellar magnetic field is frozen to the gas are
huge. The magnetic Reynolds number Re$_M$ is the ratio of the Ohmic
diffusion time to the dynamical time, and is typically of order
$10^{15}-10^{21}$. The second parameter, the ambipolar Reynolds number
Re$_{AD}$, is the ratio of the ion-neutral drift time to the
dynamical time. This number is typically many orders of magnitude
less than the first one, and can approach unity in dense molecular gas. 
These numbers
suggest that the magnetic field should be nearly perfectly frozen
to the plasma component of the gas, and generally well
frozen to the neutrals, except in the densest, nearly absolutely neutral regions.
Thus, the $B(n)$ relation is determined primarily by dynamical 
rather than by microscopic processes. Parameterizing the relation
by $q\equiv d\ln B/d \ln n$, one finds $q=1$ for compression
perpendicular to $\mbfB$, $q=2/3$ for isotropic compression,
$q=1/2$ for self-gravitating, magnetically sub-critical clouds 
\citep{1993ApJ...415..680F}, and $q=0$ for compression parallel to $\mbfB$. 

Observations of the $B(n)$ relation in molecular gas indeed show
that the strongest fields are associated with the densest gas
(e.g. \citealp{1999ApJ...520..706C}). However, in the more diffuse ISM,
the $B(n)$ relation is consistent with $q\approx 0$ over
3 orders of magnitude in $n$ (\citealp{1986ApJ...301..339T}, \citealp{2005ApJ...624..773H}). Thus, processes
beyond microscopic diffusion are required to decouple field and density.
The possibility of accelerating the decoupling through turbulent transport
has been explored by \citet{2002ApJ...567..962Z} and \citet{2004ApJ...603..165H} (see also 
\citealp{2002ApJ...578L.113K}, \citealp{2002ApJ...570..210F}, and 
\citealp{2004ApJ...609L..83L}). Numerical evidence for a weak $B(n)$ 
relation includes \citet{1999ApJ...526..279P} and \citet{2005A&A...436..585D}.

\begin{figure*}
  \includegraphics[width=\textwidth]{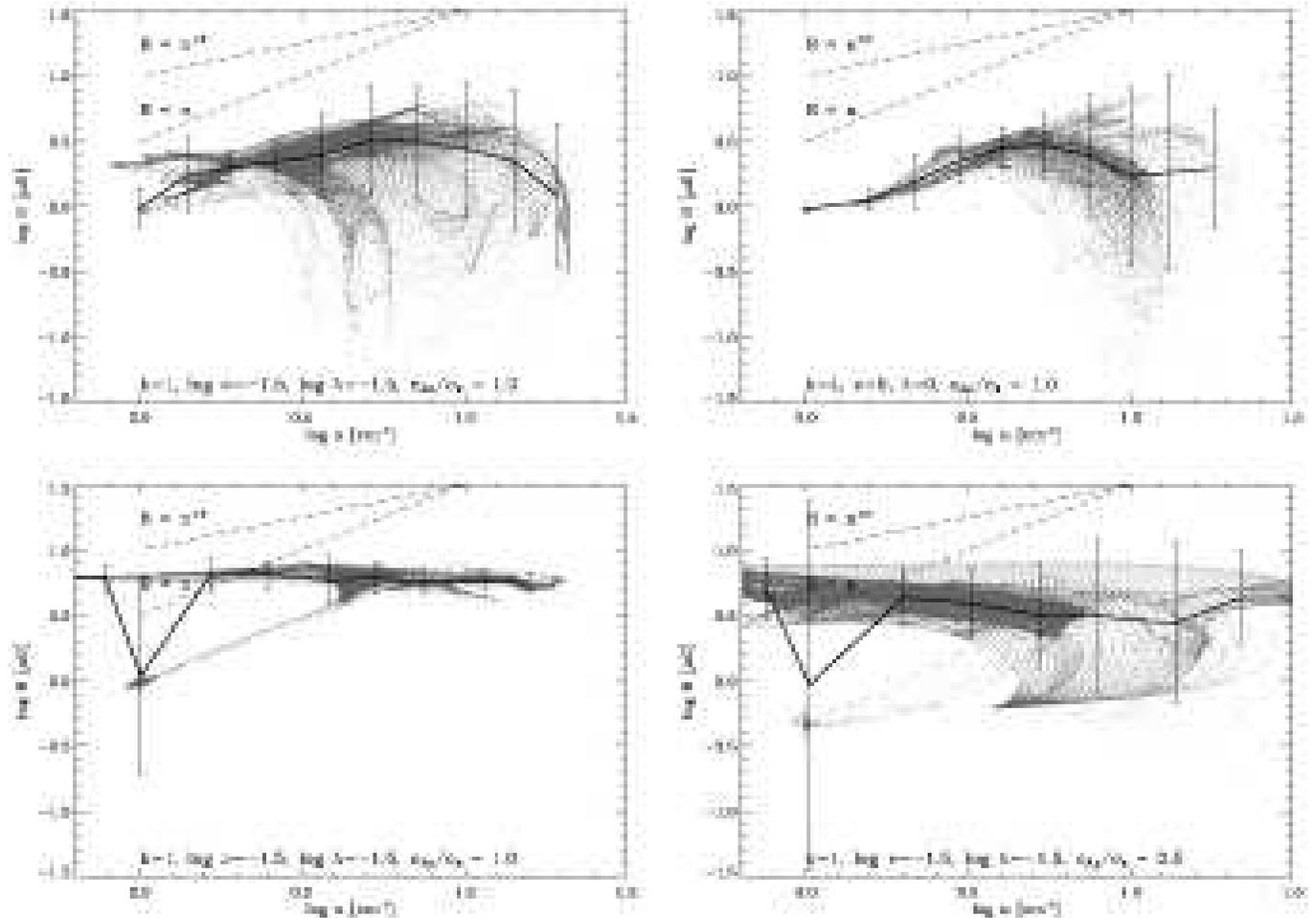}
  \caption{\label{f:fielddensity}Scatter plots of
           magnetic field strength $\log B$ against density $\log n$ for 
           models with parameters indicated in panels. Dashed lines denote idealized
           $B(n)$-scalings.}
\end{figure*}

Figure~\ref{f:fielddensity} shows scatter plots of $\log B$ against
$\log n$ for four models, each measured at $t=4$~Myr. The top row
shows models with the field parallel to the inflow (denoted by
$c_{Ax}$ in the label), while the field is oriented transversally, 
or perpendicularly to the inflow in the bottom row (denoted by
$c_{Ay}$). Idealized scalings are indicated by the dashed lines.

The most striking difference between the top and the bottom row is that
for the configuration where the field is parallel to the inflow, field and density seem to 
correlate more strongly than for the configuration where the field is perpendicular to the inflow. 
This might seem surprising at first. After all, one would expect the field
to be correlated least with density if the gas is compressed along the
field lines (top row), and to be correlated most with density when the
gas is compressed perpendicularly to the field lines (bottom row). 
However, looking back at Figures~\ref{f:emag} and \ref{f:emagy} on the one hand, 
and Figure~\ref{f:turb} on the other, we realize that the models with fields
parallel to the inflow generally develop turbulence, leading to field line 
stretching. Thus, in the (denser) slab, the field generally will be stronger.
Why, then, is there close to no correlation observable in the bottom row
of Figure~\ref{f:fielddensity}? The answer is hidden in the way we set up 
the initial conditions. To avoid the generation of strong MHD waves, we
kept the field uniform, despite the fact that the flow collision interface
is strongly perturbed (see top row of Fig.~\ref{f:denshdmhd} for the initial 
conditions). Thus, when gas is deflected at the flanks (point ``0'' in
Figure~\ref{f:sketch}), it is essentially free to move along the field lines,
i.e. transversally, but is increasingly prevented from continuing its trip towards
the slab, because the magnetic pressure is increasing (note that the bulk magnetic
field strength is higher in the bottom row of Fig.~\ref{f:fielddensity}). Thus,
the density can take on any values in the slab, while the field value is given
by the ram pressure. A weaker field (lower right panel vs lower left panel
of Figure~\ref{f:fielddensity}) reduces this effect, allowing some scatter
in the field strength, and, indeed, checking Figure~\ref{f:emagy} for this case,
the slab actually shows substructure and is allowed to bend. Note that there is
a strict $d \ln B/d\ln n=1$ scaling for the strong field model (lower left panel),
which results from the initial compression of the field lines.

While it is hard to generalize the results of our two-dimensional models, they 
demonstrate that the $B(n)$ relation is strongly influenced by the geometry of
the fields and the gas flows. Even compression perpendicular to the field lines
can lead to a nearly complete decoupling of field and density -- as long as the
gas flow is given the chance to break the symmetry.

In a sense, we expect the ``geometrical'' mechanism decoupling the field from the density 
(lower row of Figure~\ref{f:fielddensity}) to compete with the turbulent transport 
of the magnetic field: both act on dynamical (flow) timescales. For fields oriented 
perpendicularly to the inflow, our models do not develop any substantial turbulence
(this might also be due to a too small Reynolds number). On the other hand, the models with 
field parallel to the inflow direction do generate some turbulence. 
For these models, $B$ and $n$ decorrelate only at higher density values, corresponding 
to small scales, while the lower density regions show a reasonably 
well established correlation between $B$ and $n$.

%
%
\section{Summary\label{s:summary}}

The Non-linear Thin Shell Instability (NTSI, \citealp{1994ApJ...428..186V})
is expected to occur in expanding shells, shocks or colliding gas streams. 
Previous studies have addressed the evolution of the NTSI under hydrodynamical
conditions, including gravity and cooling. We have presented 
a numerical study of the NTSI including magnetic fields. We have established that our numerical 
method is well suited to tackle the problem. We have found that the effects of magnetic fields on the NTSI
can be summarized as follows:

(1) Fields principally tend to weaken or even suppress the NTSI. We further distinguish
between two cases: (i) fields aligned with the inflow resist the transverse momentum
transport -- which is the main driving agent of the NTSI -- via the magnetic
tension force; (ii) fields perpendicular to the inflow lead to a stiffer equation
of state. If $c_A\approx u$, the NTSI is suppressed. However, even for
transverse fields, substructures can form within the slab, which can serve as
fragmentation seeds in the presence of thermal instabilities or self-gravity.

(2) A fixed physical scale both for viscous and resistive dissipation is necessary
to reach numerical convergence. When relying on numerical dissipation at the resolution
scale, the Reynolds number will increase with resolution, leading to a more turbulent
environment and thus to results which qualitatively and quantitatively depend on resolution
(Figs~\ref{f:emag}, \ref{f:turb} and \ref{f:growthrates_k4}).

(3) At larger Reynolds numbers, turbulent reconnection plays a role in the turbulent
dense slab generated by the NTSI. Magnetic energy is therefore dissipated at higher rates,
leading to a pressure deficit in the dense slab. The magnetic field acts as a dissipation
channel (Figs~\ref{f:pressprof} and \ref{f:pressequi}).

(4) Although the energies (or average pressures) seem to show a tendency of the system 
to evolve towards equipartition, pressures do not balance locally within the 
slab (Figs.~\ref{f:pressequi} and \ref{f:presscorrel}). Correlated pressures 
lead to waves, i.e. the slab's inner structure is highly dynamical.

(5) The relation between field and density is, at best, weak in all models 
(Fig.~\ref{f:fielddensity}). Models with fields parallel to 
the inflow exhibit a stronger $B(n)$ correlation than models with fields
oriented perpendicularly to the inflow. The main reason for this is the
generation of turbulence, which leads to field line stretching and thus
field amplification within the denser slab. Fields oriented perpendicularly
to the inflow allow instreaming material to move laterally, permitting the 
field and density to decorrelate.

Our isothermal models only allow a limited exploration of the effect of fields
on colliding flows in a thermally or gravitationally unstable medium. Clearly,
substructure can form in the slab under most conditions, providing potential
seeds for thermal or gravitational instabilities. Thus, to establish 
the role of magnetic fields for molecular cloud formation in the colliding flow
scenario, the thermal and gravitational effects have to be addressed.

\acknowledgements
We thank E.~Zweibel for a critical reading of the manuscript and for enlightening 
discussions, and E.~Vishniac for comments on magnetic field effects in the NTSI. 
Computations were performed at the NCSA (AST040026) and on the local resources
at U of M, perfectly administered and maintained by J.~Hallum.
This work was supported by the University of Michigan
and has made use of the NASA Astrophysics Data System.

%
%

\bibliographystyle{apj}
\bibliography{./references}

\begin{thebibliography}{43}
\expandafter\ifx\csname natexlab\endcsname\relax\def\natexlab#1{#1}\fi

\bibitem[{{Balsara}(1998)}]{1998ApJS..116..133B}
{Balsara}, D.~S. 1998, \apjs, 116, 133

\bibitem[{{Bergin} {et~al.}(2004){Bergin}, {Hartmann}, {Raymond}, \&
  {Ballesteros-Paredes}}]{2004ApJ...612..921B}
{Bergin}, E.~A., {Hartmann}, L.~W., {Raymond}, J.~C., \& {Ballesteros-Paredes},
  J. 2004, \apj, 612, 921

\bibitem[{{Bhatnagar} {et~al.}(1954){Bhatnagar}, {Gross}, \&
  {Krook}}]{1954PhRv...94..511B}
{Bhatnagar}, P.~L., {Gross}, E.~P., \& {Krook}, M. 1954, Physical Review, 94,
  511

\bibitem[{{Blondin} \& {Marks}(1996)}]{1996NewA....1..235B}
{Blondin}, J.~M. \& {Marks}, B.~S. 1996, New Astronomy, 1, 235

\bibitem[{{Boldyrev} {et~al.}(2002){Boldyrev}, {Nordlund}, \&
  {Padoan}}]{2002PhRvL..89c1102B}
{Boldyrev}, S., {Nordlund}, {\AA}., \& {Padoan}, P. 2002, Physical Review
  Letters, 89, 031102

\bibitem[{{Cho} \& {Lazarian}(2003)}]{2003MNRAS.345..325C}
{Cho}, J. \& {Lazarian}, A. 2003, \mnras, 345, 325

\bibitem[{{Crutcher}(1999)}]{1999ApJ...520..706C}
{Crutcher}, R.~M. 1999, \apj, 520, 706

\bibitem[{{de Avillez} \& {Breitschwerdt}(2005)}]{2005A&A...436..585D}
{de Avillez}, M.~A. \& {Breitschwerdt}, D. 2005, \aap, 436, 585

\bibitem[{{Elmegreen} \& {Scalo}(2004)}]{2004ARA&A..42..211E}
{Elmegreen}, B.~G. \& {Scalo}, J. 2004, \araa, 42, 211

\bibitem[{{Falgarone} {et~al.}(1994){Falgarone}, {Lis}, {Phillips}, {Pouquet},
  {Porter}, \& {Woodward}}]{1994ApJ...436..728F}
{Falgarone}, E., {Lis}, D.~C., {Phillips}, T.~G., {Pouquet}, A., {Porter},
  D.~H., \& {Woodward}, P.~R. 1994, \apj, 436, 728

\bibitem[{{Fatuzzo} \& {Adams}(2002)}]{2002ApJ...570..210F}
{Fatuzzo}, M. \& {Adams}, F.~C. 2002, \apj, 570, 210

\bibitem[{{Fiedler} \& {Mouschovias}(1993)}]{1993ApJ...415..680F}
{Fiedler}, R.~A. \& {Mouschovias}, T.~C. 1993, \apj, 415, 680

\bibitem[{{Gardiner} \& {Stone}(2005)}]{2005JCoPh.205..509G}
{Gardiner}, T.~A. \& {Stone}, J.~M. 2005, Journal of Computational Physics,
  205, 509

\bibitem[{{Goldreich} \& {Sridhar}(1995)}]{1995ApJ...438..763G}
{Goldreich}, P. \& {Sridhar}, S. 1995, \apj, 438, 763

\bibitem[{{Hartmann} {et~al.}(2001){Hartmann}, {Ballesteros-Paredes}, \&
  {Bergin}}]{2001ApJ...562..852H}
{Hartmann}, L., {Ballesteros-Paredes}, J., \& {Bergin}, E.~A. 2001, \apj, 562,
  852

\bibitem[{{Hawley} \& {Stone}(1995)}]{1995CoPhyCom.89..127H}
{Hawley}, J.~F. \& {Stone}, J.~M. 1995, Comp. Phys. Comm., 89, 127

\bibitem[{{Heiles} \& {Troland}(2005)}]{2005ApJ...624..773H}
{Heiles}, C. \& {Troland}, T.~H. 2005, \apj, 624, 773

\bibitem[{{Heitsch} {et~al.}(2005){Heitsch}, {Burkert}, {Hartmann}, {Slyz}, \&
  {Devriendt}}]{2005ApJ...633L.113H}
{Heitsch}, F., {Burkert}, A., {Hartmann}, L.~W., {Slyz}, A.~D., \& {Devriendt},
  J.~E.~G. 2005, \apjl, 633, L113

\bibitem[{{Heitsch} {et~al.}(2006){Heitsch}, {Slyz}, {Devriendt}, {Hartmann},
  \& {Burkert}}]{2006ApJ...648.1052H}
{Heitsch}, F., {Slyz}, A.~D., {Devriendt}, J.~E.~G., {Hartmann}, L.~W., \&
  {Burkert}, A. 2006, \apj, 648, 1052

\bibitem[{{Heitsch} \& {Zweibel}(2003)}]{2003ApJ...590..291H}
{Heitsch}, F. \& {Zweibel}, E.~G. 2003, \apj, 590, 291

\bibitem[{{Heitsch} {et~al.}(2004){Heitsch}, {Zweibel}, {Slyz}, \&
  {Devriendt}}]{2004ApJ...603..165H}
{Heitsch}, F., {Zweibel}, E.~G., {Slyz}, A.~D., \& {Devriendt}, J.~E.~G. 2004,
  \apj, 603, 165

\bibitem[{{Hueckstaedt}(2003)}]{2003NewA....8..295H}
{Hueckstaedt}, R.~M. 2003, New Astronomy, 8, 295

\bibitem[{{Kim} \& {Diamond}(2002)}]{2002ApJ...578L.113K}
{Kim}, E.-j. \& {Diamond}, P.~H. 2002, \apjl, 578, L113

\bibitem[{{Klein} \& {Woods}(1998)}]{1998ApJ...497..777K}
{Klein}, R.~I. \& {Woods}, D.~T. 1998, \apj, 497, 777

\bibitem[{{Li} \& {Nakamura}(2004)}]{2004ApJ...609L..83L}
{Li}, Z.-Y. \& {Nakamura}, F. 2004, \apjl, 609, L83

\bibitem[{{Mouschovias} \& {Spitzer}(1976)}]{1976ApJ...210..326M}
{Mouschovias}, T.~C. \& {Spitzer}, Jr., L. 1976, \apj, 210, 326

\bibitem[{{Padoan} \& {Nordlund}(1999)}]{1999ApJ...526..279P}
{Padoan}, P. \& {Nordlund}, {\AA}. 1999, \apj, 526, 279

\bibitem[{{Palotti} {et~al.}(2006){Palotti}, {Heitsch}, {Zweibel}, \&
  {Huang}}]{palotti2006submit}
{Palotti}, M.~L., {Heitsch}, F., {Zweibel}, E.~G., \& {Huang}, Y.-M. 2006,
  \apj, submitted

\bibitem[{{Passot} {et~al.}(1995){Passot}, {V\'{a}zquez-Semadeni}, \&
  {Pouquet}}]{1995ApJ...455..536P}
{Passot}, T., {V\'{a}zquez-Semadeni}, E., \& {Pouquet}, A. 1995, \apj, 455, 536

\bibitem[{{Prendergast} \& {Xu}(1993)}]{1993JCoPh.109...53P}
{Prendergast}, K.~H. \& {Xu}, K. 1993, Journal of Computational Physics, 109,
  53

\bibitem[{{Shu} \& {Osher}(1988)}]{1988JCP..77..439S}
{Shu}, C.-W. \& {Osher}, S. 1988, \jcp, 77, 439

\bibitem[{{Shu} {et~al.}(1987){Shu}, {Adams}, \&
  {Lizano}}]{1987ARA&A..25...23S}
{Shu}, F.~H., {Adams}, F.~C., \& {Lizano}, S. 1987, \araa, 25, 23

\bibitem[{{Slyz} {et~al.}(2006){Slyz}, {Devriendt}, {Bryan}, {Heitsch}, \&
  {Silk}}]{slyz2006submit}
{Slyz}, A., {Devriendt}, J.~E.~G., {Bryan}, G.~L., {Heitsch}, F., \& {Silk}, J.
  2006, \mnras, submitted

\bibitem[{{Slyz} \& {Prendergast}(1999)}]{1999A&AS..139..199S}
{Slyz}, A. \& {Prendergast}, K.~H. 1999, \aaps, 139, 199

\bibitem[{{Tang} \& {Xu}(2000)}]{2000JCoPh.165...69T}
{Tang}, H.-Z. \& {Xu}, K. 2000, \jcp, 165, 69

\bibitem[{{T\'{o}th} \& {Odstr\v{c}il}(1996)}]{1996JCoPh.128...82T}
{T\'{o}th}, G. \& {Odstr\v{c}il}, D. 1996, \jcp, 182, 82

\bibitem[{{Troland} \& {Heiles}(1986)}]{1986ApJ...301..339T}
{Troland}, T.~H. \& {Heiles}, C. 1986, \apj, 301, 339

\bibitem[{{V\'{a}zquez-Semadeni} {et~al.}(1995){V\'{a}zquez-Semadeni},
  {Passot}, \& {Pouquet}}]{1995ApJ...441..702V}
{V\'{a}zquez-Semadeni}, E., {Passot}, T., \& {Pouquet}, A. 1995, \apj, 441, 702

\bibitem[{{V{\'a}zquez-Semadeni} {et~al.}(2006){V{\'a}zquez-Semadeni}, {Ryu},
  {Passot}, {Gonz{\'a}lez}, \& {Gazol}}]{2006ApJ...643..245V}
{V{\'a}zquez-Semadeni}, E., {Ryu}, D., {Passot}, T., {Gonz{\'a}lez}, R.~F., \&
  {Gazol}, A. 2006, \apj, 643, 245

\bibitem[{{Vishniac}(1994)}]{1994ApJ...428..186V}
{Vishniac}, E.~T. 1994, \apj, 428, 186

\bibitem[{{Xu}(1999)}]{1999JCoPh.153..334X}
{Xu}, K. 1999, \jcp, 153, 334

\bibitem[{{Zachary} {et~al.}(1994){Zachary}, {Malagoli}, \&
  {Collella}}]{1994JSSC..15..263Z}
{Zachary}, A.~L., {Malagoli}, A., \& {Collella}, P. 1994,
  J.~Sci.~Stat.~Comput., 15, 263

\bibitem[{{Zweibel}(2002)}]{2002ApJ...567..962Z}
{Zweibel}, E.~G. 2002, \apj, 567, 962

\end{thebibliography}

\end{document}